\documentclass[aps,prl,twocolumn,superscriptaddress,nofootinbib,floatfix]{revtex4-2}

\usepackage{amsmath,amssymb,mathtools}
\usepackage{graphicx}
\usepackage{xcolor}
\usepackage{hyperref}
\hypersetup{hidelinks}

\newcommand{\Mpl}{M_\mathrm{pl}}

\newcommand{\calR}{\mathcal{R}}
\newcommand{\calC}{\mathcal{C}}

\newcommand{\calN}{\mathcal{N}}
\newcommand{\calO}{\mathcal{O}}
\newcommand{\be}{\begin{equation}}
\newcommand{\ee}{\end{equation}}

\newcommand{\hD}{\hat{\Delta}}

\begin{document}

\title{Primordial black holes forming during kination:\\ the trapped, the overdense, and the void}

\author{Cristian Joana}
\email{cristian.joana@ucas.ac.cn}
\affiliation{Center for Theoretical Physics, School of Physics and Optoelectronic Engineering,\\ Hainan University, Haikou 570228, China}
\affiliation{International Centre for Theoretical Physics Asia-Pacific,
University of Chinese Academy of Sciences, 100190 Beijing, China}

\date{\today}

\begin{abstract}
We study primordial black-hole (PBH) formation in a single-field model which passes from slow roll, through a transient ultra-slow-roll phase, into kination.
Nonlinear numerical-relativity simulations are used to follow localised, initially sub-Hubble field fluctuations across the non-attractor transition, to determine the resulting super-Hubble curvature profiles, and subsequently to evolve their re-entry during kination.
Within the set of initial configurations explored here, we find three qualitatively distinct outcomes, namely a patch which remains trapped in inflation, a positive-curvature overdensity, and a negative-curvature void bounded by an overdense shell.
Each may form a PBH, although by a different physical mechanism.
For one selected profile in each of the overdense and void channels, we measure collapse thresholds and compare them with thresholds for sinc profiles.
The comparison shows that neither the central curvature nor the peak linear compaction alone provides a profile- or channel-independent collapse criterion.
PBH abundances in non-attractor models must therefore be formulated in terms of the distribution of complete profiles, with the trapped, overdense, and void channels treated separately.
\end{abstract}

\maketitle

Primordial black holes (PBHs) could comprise some or all of the dark matter~\cite{Carr:2020xqk,Green:2020jor,Carr:2020gox}, seed the supermassive black holes observed at high redshift~\cite{Carr:2018rid,Sasaki:2018dmp}, and contribute to a stochastic gravitational-wave background~\cite{NANOGrav:2023hvm,EPTA:2023fyk,Reardon:2023gzh,Xu:2023wog,LISACosmologyWorkingGroup:2023njw,Shankaranarayanan:2026pbh}.
In the usual picture, a sufficiently large primordial curvature perturbation collapses after entering the Hubble radius~\cite{Hawking:1971ei,Carr:1974nx,Carr:1975qj,Khlopov:1985fch}.
An ultra-slow-roll (USR) episode, during which the inflaton velocity drops and the comoving curvature perturbation grows outside the Hubble radius~\cite{Motohashi:2014ppa,Motohashi:2017aob,Motohashi:2019rhu,Pi:2020otn,Tomberg:2023kli}, provides an efficient mechanism of producing this enhancement.
The relation between field fluctuations and curvature is then nonlinear and can generate substantial local non-Gaussianity~\cite{Pattison:2017mbe,Ezquiaga:2019ftu,Figueroa:2020jkf,Pi:2022ysn,Caravano:2024moy,Caravano:2025diq}.

A prediction for PBH formation requires more than the amplitude of this enhancement; it depends upon the full spatial profile of the curvature perturbation and upon the background through which it re-enters.
In practice the two are treated separately.
Collapse calculations begin from a prescribed super-Hubble profile in a chosen fluid~\cite{Musco:2004ak,Musco:2012au,Harada:2015yda,Escriva:2019phb,Escriva:2020tak,Escriva:2022duf,Escriva:2025rja,Yuwen:2026bulk}, while calculations of USR statistics commonly end once the curvature has frozen and use thresholds obtained for template profiles~\cite{Atal:2019cdz,Atal:2019erb,Inui:2024fgk,Young:2022phe}.
This separation obscures the dynamical relation between the initial perturbation and the profile which eventaully collapses; indeed, to our knowledge, the complete sequence from an initially sub-Hubble field fluctuation, through its nonlinear amplification, to Hubble re-entry and black-hole formation has not previously been followed in a single continuous nonlinear calculation.
The gap is not specific to black holes formed during kination. In the model considered here, however, the final stage occurs during kination, so the scalar field must be evolved rather than replaced by radiation or an ideal stiff fluid.

In this Letter, we treat the problem end to end using numerical relativity, a well-established tool for nonlinear early-Universe dynamics~\cite{Clough:2016ymm,Clough:2017efm,Joana:2020rxm,Joana:2022uwc,Joana:2024ltg,Aurrekoetxea:2019fhr,Elley:2024alx,Brady:2025zxp,Brady:2026evi,Aurrekoetxea:2024ypv} and for PBH formation in scalar-field cosmologies~\cite{deJong:2021bbo,Kou:2019bbc,Cheng:2025eas,Padilla:2025bkv,Milligan:2025zbu,2026arXiv260421520P}. 
Working in spherical symmetry, we select three localised, initially sub-Hubble field perturbations and evolve each continuously through slow roll (SR), USR and kination to apparent-horizon formation, with only the gauge adapted during collapse and nothing prescribed or matched between stages.
Dedicated scans then compute collapse thresholds, both for frozen curvature profiles extracted from these evolutions and for analytic sinc references.
{ Throughout this work we treat the system classically within full general relativity, however we note that any effects due to the stochastic quantum nature of the field have been neglected.
}

The principal result is that a single family of initial fluctuations produces three nonlinear routes to PBH formation.
A sufficiently delayed patch remains trapped in inflation and is hidden behind a horizon by a process akin to false-vacuum-bubble collapse~\cite{Blau:1986cw,Garriga:2015fdk,Deng:2017uwc,Deng:2020pxo,Escriva:2023vacuum,Wang:2025bubble,Franciolini:2026eternity}.
A smaller delay produces the usual positive-curvature overdensity.
An advanced patch instead becomes a negative-curvature void whose compensating overdense shell may collapse and drive the core through an overdense rebound~\cite{Joana:2025gqf}.
{For the overdense case, the critical central amplitudes of the generated and sinc profiles differ significantly by about $88\%$, while their peak linear compactions differ by about $19\%$.}
For a void, it is the nonlinear dynamics after the formation of the surrounding shell, rather than the central curvature, which governs collapse~\cite{Joana:2025gqf}.

For the same sinc profiles, we also find thresholds in scalar-field kination below those for an ideal stiff fluid.
These results show that PBH predictions in this class of non-attractor model require profile-resolved statistics and separate accounting of the trapped, overdense, and void channels.

\begin{figure*}[p]
    \centering
    \includegraphics[width=\textwidth]{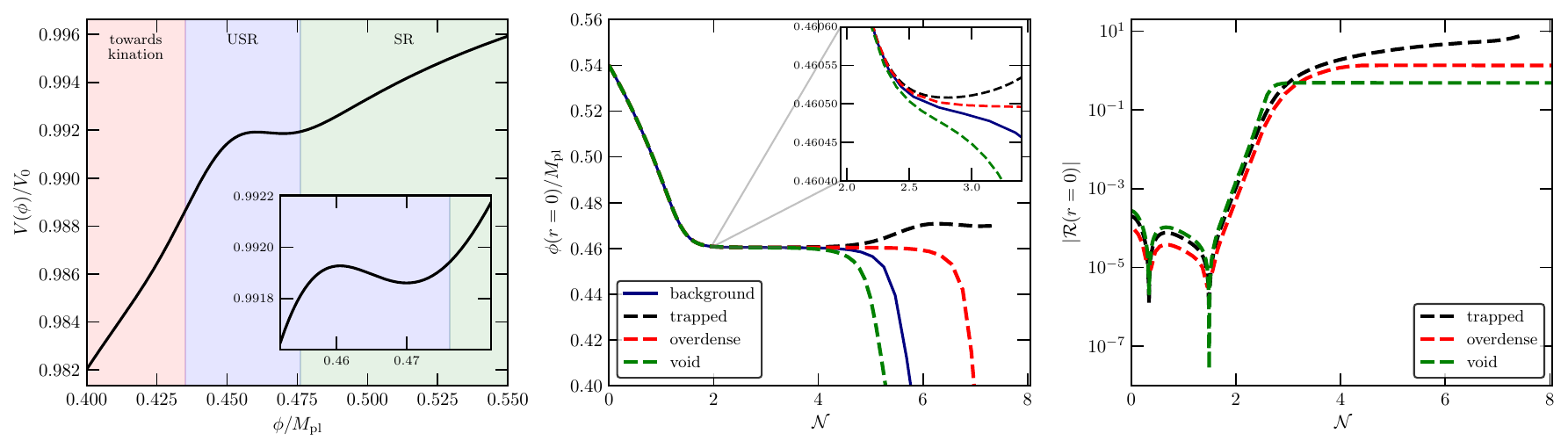}
    \caption{%
    {The left panel shows the scalar potential Eq.~\eqref{eq:potential}, with the SR (green), USR (blue), and evolution towards kination (red) portions of the background evolution shaded; the inset resolves the shallow minimum at $\phi\simeq0.470\,\Mpl$ which supports the trapped solution.}
    The centre panel shows the central field value $\phi(r{=}0)$ (dashed) and homogeneous background $\overline\phi$ (solid) against e-fold number $\calN$ for the trapped (black), overdense (red), and void (green) initial data.
    The right panel shows the corresponding central curvature $|\calR(r{=}0)|$ and its nonlinear amplification during USR.
    }
    \label{fig:dynamics}
\end{figure*}

\textit{Inflationary model.---}%
We take a minimally coupled scalar field $\phi$ with a T-model plateau~\cite{Martin:2013tda,Kallosh:2013hoa} modified by a Gaussian feature,
\be
V(\phi) = \frac{V_0}{2}\!\left[1 + \tanh\!\left(\frac{\phi}{\varphi}\right)\right] + V_0\, \lambda \exp\!\left[-\frac{(\phi - \phi_0)^2}{2\sigma^2}\right] ,
\label{eq:potential}
\ee
{where $V_0 = 8.0727656\cdot 10^{-13}\, \Mpl^4$, $\varphi = 0.2\,\Mpl$, $\phi_0 = 0.45\,\Mpl$, $\sigma = 0.014\,\Mpl$, and $\lambda = 2.4250043\cdot 10^{-3}$.}
The hyperbolic tangent produces a SR plateau together with a transition to kination as $V\to0$, whilst the Gaussian bump locally slows the field and induces a transient USR phase.
{Their superposition produces a shallow local minimum at $\phi\simeq0.470\,\Mpl$, shown in the inset of Fig.~\ref{fig:dynamics}, in which sufficiently delayed field configurations may become classically trapped.}
The homogeneous background accordingly passes through SR, USR and kination, with the equation of state approaching $w=1$.
{The parameters are chosen to reproduce the observed CMB amplitude 60 e-folds before the end of inflation and to generate a small-scale peak of amplitude $\mathcal{A}_\zeta\simeq0.01$ (see the Supplemental Material).}

We solve the Einstein--Klein--Gordon system in spherical symmetry with a Z4c/BSSN formulation~\cite{Baumgarte_1998,PhysRevD.52.5428,Alic:2011gg,Bernuzzi:2009ex}.
The initial field is a localised fluctuation about the homogeneous inflationary solution,
\be
\phi(r,0) = \overline{\phi}(0) + \phi_{\rm amp}\, \mathrm{sinc}(k_\star r)\,W(r) \,,
\label{eq:IC}
\ee
where $\overline{\phi}(0)$ is the homogeneous Friedmann--Lema\^itre--Robertson--Walker (FLRW) value, $\phi_{\rm amp}$ is the amplitude, and $k_\star$ fixes the comoving scale.
The window $W(r)$ is unity over the perturbation and tends smoothly to zero before the outer boundary.
We use the homogeneous SR field momentum; the momentum constraint then determines the radial expansion, and the Hamiltonian constraint fixes the conformal factor (see the Supplemental Material).

\textit{Inflationary dynamics and curvature profiles.---}%
To connect the inflationary calculation to the subsequent collapse, we evaluate the comoving curvature perturbation $\calR$ using the separate-universe expression
\begin{align}
  \calR \;&=\; \zeta_{\rm vol} \;-\; \int_{\overline\phi}^{\phi}\frac{\overline H}{\overline\Pi}\,d\varphi
  \;+\;\calO\!\left(k_\star^2/a^2H^2\right) ,
  \label{eq:Rdecomp}\\
   \;&\simeq\;  \zeta_{\rm vol} \;-\; \frac{\overline H\,\delta\phi}{\overline\Pi}  \equiv \calR_{\rm lin} \, ,
  \label{eq:Rlin}
\end{align}
where $\zeta_{\rm vol}$ is the perturbation in the logarithmic local volume, $\Pi$ is the scalar-field momentum measured by normal observers, $H$ is the local expansion rate, $\delta\phi=\phi-\overline\phi$, and an overbar denotes the background.
Equation~\eqref{eq:Rdecomp}, whose integral runs along the background trajectory, retains the nonlinear dependence on the field displacement and is the definition used below; its linearisation, Eq.~\eqref{eq:Rlin}, agrees with it on super-Hubble scales after the transition to kination.
Both the slicing and the separate-universe construction are described in the Supplemental Material.

On the rolling branch $\overline\Pi<0$.
At sufficiently early times, whilst $\delta\phi$ remains small, $\calR$ inherits the sinc shape of Eq.~\eqref{eq:IC} with the same sign as $\phi_{\rm amp}$, and oscillates while the perturbation remains sub-Hubble.
We choose $k_\star$ such that Hubble exit occurs close to the SR--USR transition, where the nonlinear amplification is strongest.
For solutions which leave the non-attractor regime, $\calR$ then settles to a conserved, model-dependent profile.
Its sign and radial form are determined by the amplitude, phase, and shape of the initial field perturbation as they are processed by the nonlinear evolution through the SR--USR--kination transition.
As the field approaches the kination attractor, the conserved curvature is progressively carried by $\zeta_{\rm vol}$.

Which of the three channels is realised depends upon the sign and magnitude of the field displacement at the SR--USR transition, where the example configurations have about $|\calR|\sim10^{-3}$.
A sufficiently large positive displacement, corresponding to a locally delayed field, traps the central inflating region whilst its surroundings enter kination.
The contrast between interior and exterior continues to grow, and the region becomes hidden behind a black-hole horizon in the manner of vacuum-bubble PBH formation~\cite{Blau:1986cw,Garriga:2015fdk,Deng:2017uwc,Deng:2020pxo}.
A more moderate positive displacement produces a frozen profile with positive central curvature, which collapses as an overdensity if it lies above the kination threshold.
A negative displacement, corresponding to a locally advanced field, produces a frozen profile with negative central curvature and a compensating overdense shell.
All three example configurations form PBHs.
The overdense and void channels do so after Hubble re-entry, while the trapped channel follows the distinct bubble-like mechanism described above.

\begin{figure*}[p]
    \centering
    \includegraphics[width=\textwidth]{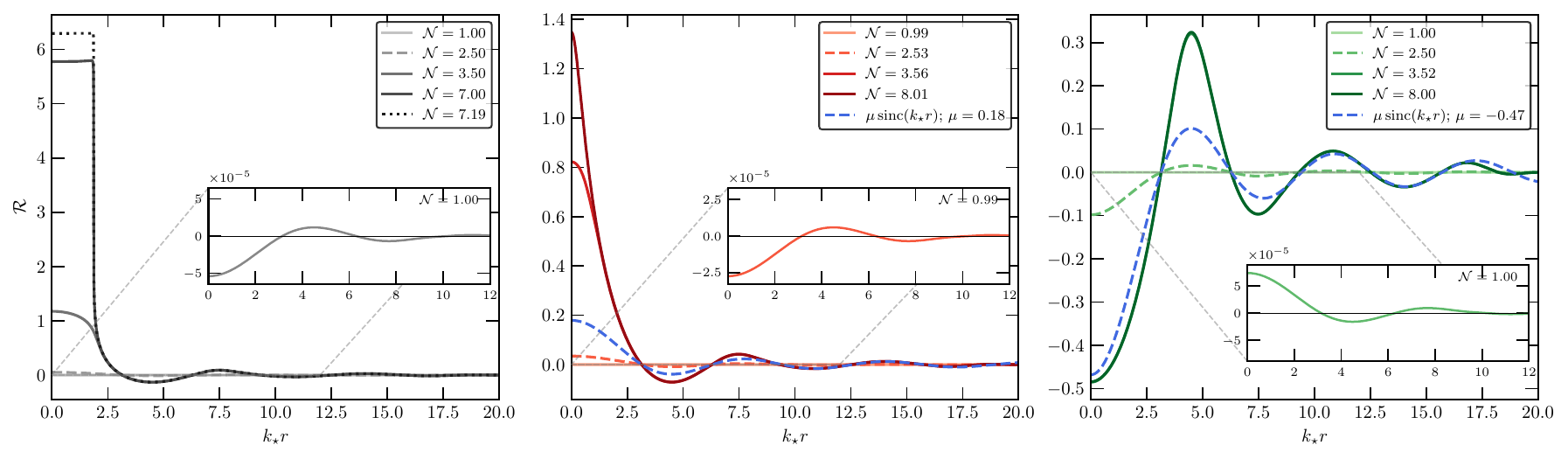}
    \caption{%
    Comoving-curvature profiles at successive e-fold numbers for the trapped (left, black), overdense (centre, red), and void (right, green) configurations; darker curves denote later times.
    The early changes of sign are sub-Hubble oscillations.
    Hubble exit is chosen to occur close to the SR--USR transition.
    Blue dashed curves are sinc profiles chosen to match the oscillatory tails; the late-time cores therefore show directly the model-dependent shape deviation from the analytic form.
    }
    \label{fig:curvature}
\end{figure*}

\begin{figure*}[p]
    \centering
    \begin{minipage}[c]{0.52\textwidth}
        \centering
        \includegraphics[width=\linewidth]{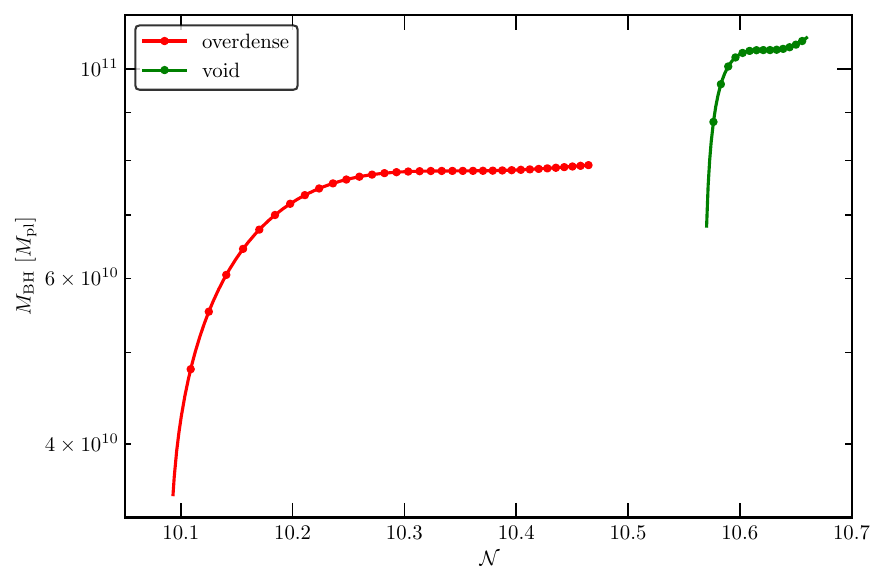}
    \end{minipage}\hfill
    \begin{minipage}[t]{0.44\textwidth}
        \caption{%
        Apparent-horizon mass $M_\mathrm{BH}$, in units of $\Mpl$, against e-fold number $\calN$ for the overdense and void configurations.
        The void remnant forms later and with a larger mass; both horizons are still accreting at the end of the simulations.}
        \label{fig:masses}
    \end{minipage}
\end{figure*}

{Figure~\ref{fig:dynamics} summarises the inflationary evolution of three illustrative configurations, initialised at $\overline\phi(0)=0.54\,\Mpl$ and $k_\star=3.3\cdot10^{-6}\,\Mpl$ with $\phi_{\rm amp}/\Mpl=+8\cdot10^{-6}$ (trapped), $+4\cdot10^{-6}$ (overdense), and $-1.1\cdot10^{-5}$ (void).}
They are illustrative rather than representative, and other amplitudes, phases, scales, or radial forms may alter both the channel and the frozen curvature profile.
{For the phase and scale chosen here, the positive initial amplitudes lead to locally delayed configurations at the SR--USR transition, whereas the negative initial amplitude leads to a locally advanced one (Fig.~\ref{fig:curvature}).}
For the trapped configuration $|\calR(r{=}0)|$ grows without approaching a stationary value, while for the overdense and void configurations it freezes on super-Hubble scales (right panel of Fig.~\ref{fig:dynamics}).
{That the trapped configuration is genuinely stalled is confirmed by the central field settling at $\phi(r{=}0)\simeq0.470\,\Mpl$, the location of the local minimum (middle panel). For the trapped channel, stochastic kicks and quantum-tunneling might my significantly affect the dynamics; however we neglect both effects in this analysis. }
Figure~\ref{fig:curvature} gives the corresponding spatial profiles $\calR(r)$ at a sequence of times.
The frozen overdense and void profiles differ noticeably from the initial sinc shape, and from one another, which motivates calibrating the collapse thresholds for the extracted profiles themselves.
Figure~\ref{fig:masses} shows the PBH masses obtained for the overdense and void configurations following re-entry, as discussed below.
For the trapped channel, the relevant quantity is instead the initial horizon mass when the locally inflating patch becomes hidden, rather than a mass inferred from an ordinary re-entry calculation~\cite{Deng:2017uwc,Deng:2020pxo,Franciolini:2026eternity}.

\begin{figure*}[!t]
    \centering
    \includegraphics[width=0.82\textwidth]{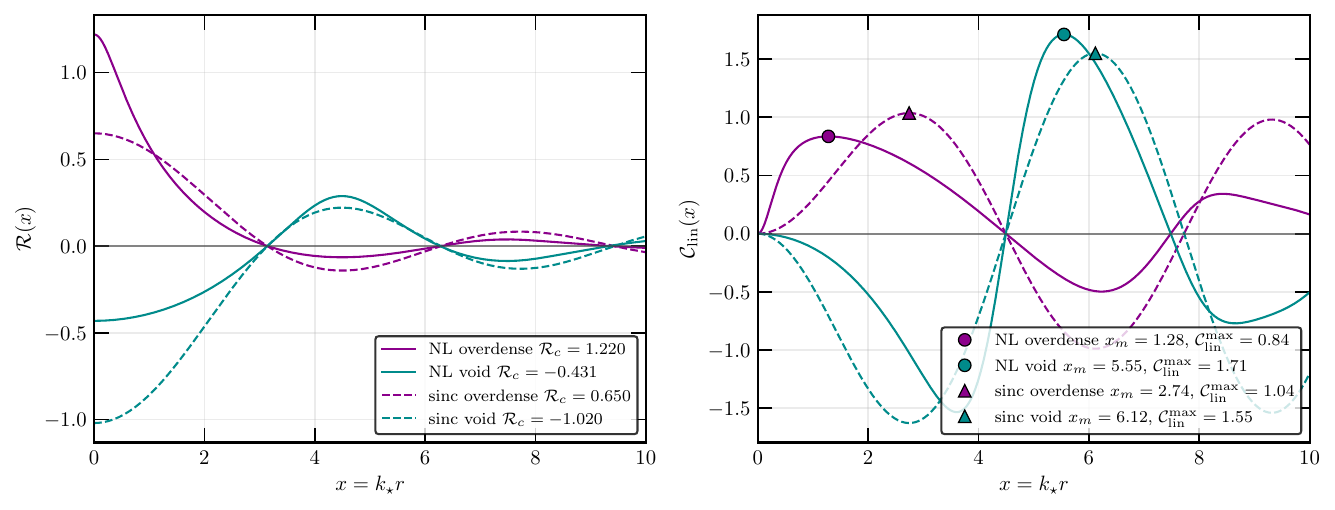}
    \caption{%
    Critical curvature profiles (left) and their linear compaction functions (right).
    Solid curves are the nonlinear (NL) profiles extracted after the SR--USR--kination evolution; dashed curves are sinc profiles at threshold.
    The markers locate the positive maximum of $\calC_{\rm lin}$ at $x_m=k_\star r_m$.
    }
    \label{fig:profile_compaction}
\end{figure*}

\textit{Collapse thresholds and PBH masses.---}%
The three solutions above are evolved continuously to apparent-horizon formation, adjusting only the gauge during collapse.
The threshold mesurements in Table~\ref{tab:thresholds_compaction}, however, come from dedicated scans.
For the overdense and void cases we extract $\calR$ after the transition to kination, once it is conserved on super-Hubble scales.
We write these profiles as
\be
    \calR(r)=\calR_0\,P(k_\star r),\qquad P(0)=1,
\ee
where $\calR_0=\calR(0)$ is the central amplitude scanned in the collapse simulations.
The same procedure is repeated for the analytic reference profile $P(x)=\mathrm{sinc}(x)$.
This benchmark is motivated by the nonlinear statistics of USR perturbations, often summarised by the local logarithmic map $\zeta=-\gamma_{\rm log}^{-1}\ln\!\left|1-\gamma_{\rm log}\zeta_G\right|$, with $f_\mathrm{NL}=\frac{5}{6}\gamma_{\rm log}$~\cite{Pi:2022ysn,Inui:2024fgk,Franciolini:2026eternity}, for which the mean spherical profile of the monochromatic Gaussian component $\zeta_G$ is proportional to a sinc function.
It therefore provides a controlled analytic reference for the threshold calculation.
In contrast, the extracted profiles used here are generated fully relativistically by the continuous SR--USR--kination evolution.
For the threshold scans we construct new, constraint-satisfying kination data with a homogeneous scalar-field background and $\Pi$ computed from the Hamiltonian constraint.
The tabulated thresholds are therefore conditional on a fixed shape $P(x)$ whose normalisation is varied, rather than thresholds along the original $\phi_{\rm amp}$ family, for which the nonlinear evolution changes both amplitude and shape.

We identify a black hole by the outermost future marginal surface, $\Theta_+=0$ and $\Theta_-<0$.
For comparison between profiles, Table~\ref{tab:thresholds_compaction} also gives the linear compaction
\begin{align}
    \calC_{\rm lin}(x) &= -2 f(w)\,x\,\partial_x \calR, \label{eq:Clin}\\
    f(w) &= \frac{3(1+w)}{5+3w},\qquad x=k_\star r ,
\end{align}
evaluated at $w=1$.
Collapse is sensitive to the radial distribution of curvature---and in particular to the surrounding shell---rather than to $\calR(0)$ alone~\cite{Polnarev:2006aa,Musco:2018rwt,Germani:2025hcu,Escriva:2025rja}.
On the homogeneous-field collapse slice $\delta\phi=0$, and the metric curvature used to initialise the simulations is $\calR$; the long-wavelength expression may therefore be applied directly.
Figure~\ref{fig:profile_compaction} compares the critical nonlinear (NL) and sinc profiles together with their linear compaction functions; the transition-generated transformation shifts the thresholds substantially.
{For the overdensity, the critical central amplitude increases from $0.65$ for the sinc profile to the adopted value $1.220$ for the extracted profile, a shift of about $88\%$, even though the peak linear compaction is about $19\%$ smaller.}
{For the void, the magnitude of central threshold is $-0.432$, compared with $-1.02$ for the sinc profile, a decrease of about $58\%$ in magnitude, whilst the peak linear compaction is about $11\%$ larger.}
The compensating shell, rather than the central value, therefore carries the collapse information in the void channel, consistent with the profile classification of Ref.~\cite{Germani:2025hcu} and the explicit collapse mechanism of Ref.~\cite{Joana:2025gqf}.
\begin{table}[t]
\caption{
Critical central amplitudes and linear-compaction diagnostics for the profiles used in the collapse simulations.
}
\label{tab:thresholds_compaction}
\begin{ruledtabular}
\begin{tabular}{lccc}
Profile & $\calR_c$ & $x_m$ & $\calC_{\rm lin}^{\rm max}$ \\
\hline
{NL overdense} & {$1.220$} & {$1.280$} & {$0.836$} \\
{NL void} & {$-0.432$} & {$5.548$} & {$1.716$} \\
sinc overdense & $0.65\pm0.01$ & $2.743$ & $1.037$ \\
sinc void & $-1.02\pm0.01$ & {$6.117$} & $1.550$ \\
\end{tabular}
\end{ruledtabular}
\end{table}
The comparison shows that peak linear compaction provides a more stable first diagnostic of the collapse threshold than the central curvature.
{Within each channel, replacing the sinc profile by the extracted shape changes $\calC_{\rm lin}^{\rm max}$ by about $19\%$ for the overdensity and $11\%$ for the void, compared with shifts of $88\%$ and $58\%$ in the magnitude of $\calR_c$.}
This does not make $\calC_{\rm lin}^{\rm max}$ universal, since its critical value remains dependent upon the profile, channel, and background~\cite{Musco:2018rwt,Germani:2023ojx,Germani:2025hcu,Escriva:2025rja}.
In an ideal stiff fluid, the overdensity and void thresholds $\calR_c^{(+)}\simeq0.72$ and $\calR_c^{(-)}\simeq-1.15$ correspond to $\calC_{\rm lin}^{\rm max}\simeq1.148$ and $1.748$, respectively~\cite{Joana:2025gqf}.
The scalar-field kination values $\calC_{\rm lin}^{\rm max}\simeq1.037$ and $1.550$ in Table~\ref{tab:thresholds_compaction} lie approximately $10\%$ and $11\%$ below their ideal-stiff counterparts.
The reduction therefore persists when the threshold is expressed in terms of peak compaction rather than central curvature.
I attribute it to the anisotropic scalar-gradient stress, whose averaged pressure $p_\mathrm{grad}=-\frac13\rho_\mathrm{grad}$ softens the effective response once gradients become relevant.

The overdense and void channels form apparent horizons containing a few per cent of the Hubble mass and subsequently undergo appreciable environment-fed accretion, as shown in Fig.~\ref{fig:masses}.
Secondary and tertiary shells continue to reach the central region during the simulations, showing that the early PBH mass depends upon the surrounding profile as well as the first collapsing peak.
These evolutions establish the formation mechanism and early growth history, whilst asymptotic masses and the critical law $M_\mathrm{BH}\propto(\calR_0-\calR_c)^\gamma$ require longer near-threshold runs.

\textit{Discussion.---}%
We have followed, in full general relativity, the amplification and collapse of a localised sub-Hubble inflaton fluctuation through SR, USR and kination to apparent-horizon formation, without prescribing or matching an intermediate curvature profile.
This end-to-end evolution, to our knowledge the first of its kind, reveals three physically distinct routes to PBH formation within a single non-attractor model.
They are not merely different signs or amplitudes of the same collapse problem; nor are they special to kination, since the same three channels may arise in other non-attractor models, although the thresholds computed here apply to collapse during kination.

In the trapped channel, a sufficiently delayed patch fails to leave inflation, and its subsequent evolution within the decelerating kination background realises a vacuum-bubble collapse~\cite{Deng:2017uwc,Deng:2020pxo,Escriva:2023vacuum,Wang:2025bubble,Ning:2026fopt,Franciolini:2026eternity}.
The relevant initial mass is the horizon mass when the locally inflating patch becomes hidden, and the abundance is controlled by the probability of trapping; ordinary near-threshold overdensity scaling is not the appropriate description.

The overdense channel proceeds through the conventional collapse mechanism, but with a profile generated by the nonlinear evolution of an initially sub-Hubble fluctuation rather than prescribed.
{The $88\%$ shift in the central-amplitude threshold, and a correspondng change of about $19\%$ in peak linear compaction, directly measures the collapse-level importance of this profile deformation.}

For the void channel, the dynamically relevant structure is the compensating ridge.
The relativly rapid expansion of the underdense core evacuates the interior and sharpens the surrounding overdensity, resembling the tendency of cosmological voids to acquire dense, approximately spherical walls~\cite{Sheth:2004void}.
At re-entry, the low pressure in the core gives an inward pressure gradient on the inner face of the ridge.
The imploding shell compresses the core into an overdense rebound, which forms a PBH if it becomes sufficiently compact~\cite{Joana:2025gqf}.
{The $58\%$ shift in the magnitude of the central threshold, with an $11\%$ change in peak linear compaction, exposes the central role of the shell geometry.}
The ridge profile is therefore part of the threshold data, in accord with the open-, flat-, and closed-core classification of Ref.~\cite{Germani:2025hcu}.

These values characterise the selected profiles. Their significance is that the generated profiles do not inherit the collapse criteria of analytic sinc profiles.
The wider implication is that a one-point distribution for $\calR(0)$ is insufficient in this class of model.
The nonlinear evolution maps each initial field fluctuation into a conserved curvature profile, determining both its central amplitude and radial shell structure.
I argue that the abundance problem should instead be posed as a distribution over profiles, with separate conditions for trapping, overdense collapse, and void-shell collapse~\cite{Young:2024jsu,Escriva:2026hel}.

PBH formation during kination also bears on reheating.
In runaway potentials the inflaton never oscillates about a minimum, so parametric and tachyonic preheating are unavailable and reheating must proceed by other means.
Since black holes redshift as matter whilst the kination background dilutes as $a^{-6}$, even a small abundance formed through these channels can grow towards domination, and its Hawking evaporation could reheat the Universe~\cite{Domenech:2024wao}, a gravitational counterpart to the preheating instablities of oscillating backgrounds~\cite{Martin:2019nuw,Auclair:2020csm}, with induced gravitational-wave signatures of its own~\cite{Domenech:2020ssp,Balaji:2024hpu,Papanikolaou:2024kjb,Bhaumik:2026rzf}. 

These results define a concrete route towards a PBH abundance calculation for non-attractor models.
The next step should be to determine the probability distribution of frozen profiles and map the three collapse surfaces within that profile space~\cite{Animali:2025pyf,Animali:2026rvi}.
Longer evolutions will separate critical scaling from environmental accretion, whilst three-dimensional simulations should determine the stability of the trapped, overdense, and void-shell channels beyond spherical symmetry.
The distinct nonlinear histories also motivate corresponding calculations of their gravitational-wave signals for LISA~\cite{LISA:2017pwj,LISACosmologyWorkingGroup:2022jok,LISACosmologyWorkingGroup:2023njw,Joana:2025bfy}, Taiji~\cite{Ruan:2018tsw}, TianQin~\cite{TianQin:2015yph}, the Einstein Telescope~\cite{Punturo:2010zz}, and pulsar timing arrays~\cite{NANOGrav:2023hvm,EPTA:2023fyk}.

\begin{acknowledgments}
The author thanks Diego Cruces, Sébastien Clesse, Shi Pi, Jun Zhang and Hardi Veerm\"ae for stimulating conversations.
C.J. is supported by NSFC Grants No. W2433007, No. E414660101, and No. 12547104, and by the Fundamental Research Funds for the Central Universities under Grants No. E4EQ6604X2 and No. E3ER6601A2.
\end{acknowledgments}

\bibliography{biblio}

\onecolumngrid
\appendix
\clearpage
\section{Supplemental Material on the Numerical Relativity Simulations}
\label{appendix:NR_spherical}

\subsection{Inflation model}

The potential is given in Eq.~\eqref{eq:potential}.
Its hyperbolic-tangent part has a slow-roll plateau at $\phi>0$ and tends to zero as the field passes $\phi=0$.
The subsequent evolution is kinetically dominated, with $w\to1$.
{The Gaussian feature, centred at $\phi_0=0.45\,\Mpl$ and having width $\sigma=0.014\,\Mpl$ and relative amplitude $\lambda=2.4250043\cdot10^{-3}$, briefly decelerates the field and produces USR.}

We obtain the linear comoving-curvature spectrum by solving the Mukhanov--Sasaki equation~\cite{Mukhanov:1985rz,Sasaki:1986hm}; Ref.~\cite{Karam:2022nym} reviews the method.
{We choose the potential parameters to satisfy the CMB constraints. In particular, the normalisation $V_0$ is fixed by requiring $\mathcal P_{\mathcal R}(k_p)=\mathcal A_s=2.101\cdot10^{-9}$ at the comoving CMB pivot $k_p=0.05\,\mathrm{Mpc}^{-1}$. We set the Hubble exit of this pivot 60 e-folds before the end of inflation; the resulting spectral index is $n_s\simeq0.9642$, consistent with the Planck constraints.}
{The feature gives a peak $\mathcal{A}_\zeta\simeq0.01$ at $k/k_p\simeq1.9\cdot10^{24}$, before the end of inflation and the onset of kination. This model produces PBHs that are expected have evaporated, and potentially serve as a reheating mechanism to radiation domination.}

\begin{figure}[h]
    \centering
    \includegraphics[width=0.8\columnwidth]{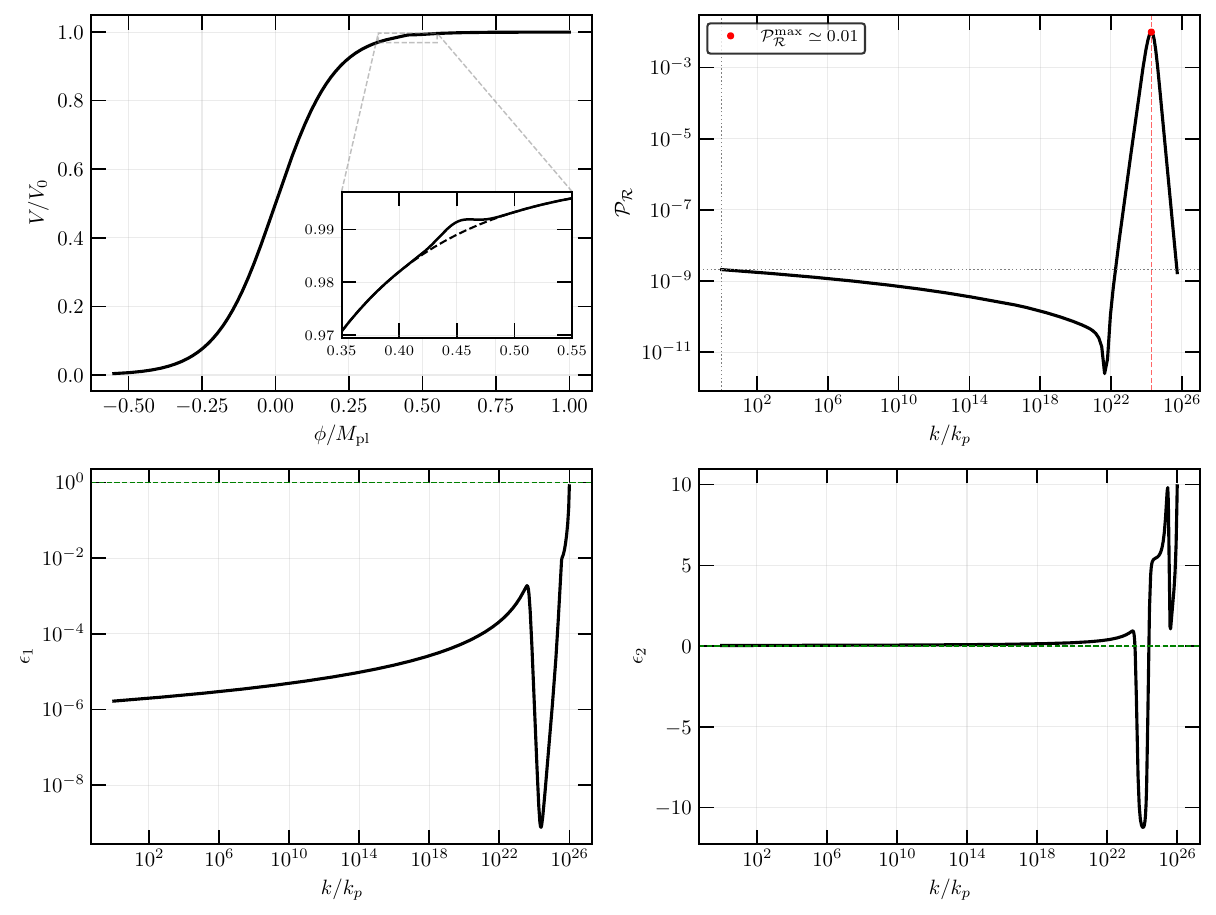}
    \caption{%
    Inflationary potential (upper left), linear curvature spectrum (upper right), and the first two Hubble-flow parameters (lower panels), with $\Mpl=1$.
    }
    \label{fig:pot_ap}
\end{figure}

The potential, spectrum, and Hubble-flow parameters $\epsilon_1$ and $\epsilon_2$ are shown in Fig.~\ref{fig:pot_ap}.

\subsection{Evolution equations}

We solve the Einstein--Klein--Gordon equations in spherical symmetry, using the BSSN implementation of Ref.~\cite{Alcubierre:2011pkc} and its Z4c extension~\cite{Alic:2011gg,Bernuzzi:2009ex}.
Reduced Planck units are used, $\Mpl=1$, and hence $8\pi G=1$.

In the ADM $3{+}1$ decomposition the line element is
\be
ds^2 = -\alpha^2\,dt^2 + \gamma_{ij}\!\left(dx^i + \beta^i\,dt\right)\!\left(dx^j + \beta^j\,dt\right) ,
\ee
where $\alpha$, $\beta^i$, and $\gamma_{ij}$ are the lapse, shift, and spatial metric.
In spherical symmetry we write
\be
dl^2 = e^{4\chi}\!\left[a(r,t)\,dr^2 + r^2\, b(r,t)\,d\Omega^2\right] ,
\ee
where $\chi$ is the conformal factor and $a$ and $b$ are conformal-metric components.
For the trace-free extrinsic curvature we define $\tilde{A}^r{}_r=A_a$ and $\tilde{A}^\theta{}_\theta=A_b=-A_a/2$; $\hD^r$ denotes the conformal connection vector.
Z4c adds the constraint-damping field $\Theta$ and evolves $\widehat K=K_{\rm phys}-2\Theta$.
The physical volume expansion is therefore governed by $K_{\rm phys}=\widehat K+2\Theta$, with $H=-K_{\rm phys}/3$.
The choice $\kappa_1=0$ and $\Theta=0$ gives BSSN.

The geometrical variables $\{\chi,a,b,\widehat K,A_a,\hD^r,\Theta,\alpha,\beta^r\}$ obey~\cite{Alcubierre:2011pkc,Bernuzzi:2009ex}
\begin{eqnarray}
\partial_t \chi &=& \beta^r \partial_r \chi
- \frac{1}{6} \alpha \left(\widehat K + 2\Theta\right) , \\
\partial_t a &=& \beta^r \partial_r a + 2 a \partial_r \beta^r
- 2 \alpha a A_a , \qquad \\
\partial_t b &=& \beta^r \partial_r b + 2 b \: \frac{\beta^r}{r}
- 2 \alpha b A_b \; , \\
\partial_t \widehat K &=& \beta^r \partial_r \widehat K - \nabla^2 \alpha
+ \alpha \left[ A_a^2 + 2 A_b^2 + \frac{1}{3} \left(\widehat K+2\Theta\right)^2\right] \nonumber  \label{eq:CCZ4_K} \\
&+& \frac{\alpha}{2} \left( \rho_E + S_a + 2 S_b \right)
+ \kappa_1(1-\kappa_2)\,\alpha\,\Theta ,
\\
\partial_t A_a &=& \beta^r \partial_r A_a - \left( \nabla^r \nabla_r \alpha
- \frac{1}{3} \nabla^2 \alpha \right)  \nonumber \\
&+& \alpha \left( {}^{(3)}R^r{}_r - \frac{1}{3}\, {}^{(3)}R \right) \nonumber \\
&+& \alpha \left(\widehat K+2\Theta\right) A_a - \frac{2}{3} \alpha \left( S_a - S_b \right) \; ,
\end{eqnarray}
\begin{eqnarray}
\partial_t \hD^r &=& \beta^r \partial_r \hD^r - \hD^r \partial_r \beta^r
+ \frac{1}{a} \partial^2_r \beta^r + \frac{2}{b} \:
\partial_r \left( \frac{\beta^r}{r} \right)
\nonumber \\ &-&
\frac{2}{a} \left( A_a \partial_r \alpha
+ \alpha \partial_r A_a \right)  \nonumber \\
&+& 2 \alpha \left( A_a \hD^r - \frac{2}{rb}
\left( A_a - A_b \right) \right)
\nonumber \\ &+&
\frac{2\alpha}{a} \left[ \partial_r A_a
- \frac{2}{3} \: \partial_r \widehat K  + 6 A_a \partial_r \chi
\right. \nonumber \\
&+&  \left. \left( A_a - A_b \right) \left( \frac{2}{r}
+ \: \frac{\partial_r b}{b} \right)
- S_r \right]
- \frac{2\alpha}{3a}\,\partial_r\Theta
-2\alpha\kappa_1\mathcal C_{\Delta}^{r} \;  , \label{eq:CCZ4_Deltar}
\end{eqnarray}
and the constraint-damping field obeys
\begin{eqnarray}
\partial_t \Theta &=& \beta^r \partial_r \Theta
+ \frac{\alpha}{2}\left[{}^{(3)}R - \left(A_a^2 + 2 A_b^2\right)
+ \frac{2}{3}\,\left(\widehat K+2\Theta\right)^2 \right. \nonumber \\
&& \left. - 2\rho_E \right]
- \alpha\,\Theta\,\kappa_1\!\left(2 + \kappa_2\right) . \label{eq:CCZ4_Theta}
\end{eqnarray}
Here $\mathcal C_{\Delta}^{r}=\hD^r-\hD^r_{\rm metric}$ is the conformal-connection constraint.
Terms containing $\Theta$, $\kappa_1$, and $\kappa_2$ comprise the Z4c extension~\cite{Bernuzzi:2009ex,Weyhausen:2011cg}; the threshold tests include the BSSN limit and Z4c runs with fixed $\kappa_1=10^{-4}$.
The damping scale $\kappa_1$ has dimensions of inverse time and is held fixed under refinement.
Scaling it as $1/\Delta t$ would alter the continuum problem.

The sources $\rho_E$, $S_r$, $S_a$, and $S_b$ are defined below.
The non-vanishing conformal inverse-metric components are $\tilde\gamma^{rr}=a^{-1}$, $\tilde\gamma^{\theta\theta}=1/(br^2)$, and $\tilde\gamma^{\phi\phi}=1/(br^2\sin^2\theta)$.
The required lapse derivatives are
\begin{eqnarray}
\nabla^2 \alpha &=&  \frac{1}{a e^{4 \chi}} \left[ \partial_r^2 \alpha
- \partial_r \alpha \left( \frac{\partial_r a}{2a}
- \frac{\partial_r b}{b}
- 2 \partial_r \chi - \frac{2}{r} \right) \right],  \nonumber
\\
\nabla^r \nabla_r \alpha &=& \frac{1}{a e^{4 \chi}} \left[ \partial_r^2 \alpha
- \partial_r \alpha \left( \frac{\partial_r a}{2a}
+ 2 \partial_r \chi \right) \right] \; .
\end{eqnarray}

The radial diagonal component of the three-Ricci tensor and its trace are
\begin{align}
{}^{(3)}R^r{}_r = &- \frac{1}{a e^{4 \chi}} \biggl[ \frac{\partial^2_r a}{2a}
- a \partial_r \hD^r - \frac{3}{4} \left( \frac{\partial_r a}{a} \right)^2
+ \frac{1}{2} \left( \frac{\partial_r b}{b} \right)^2 \nonumber \\
&- \frac{1}{2} \hD^r \partial_r a + \frac{\partial_r a}{rb}
+ \frac{2}{r^2} \left( 1 - \frac{a}{b} \right) \left( 1 + \frac{r \partial_r b}{b} \right) \nonumber \\
&+ 4 \partial^2_r \chi - 2 \partial_r \chi \left( \frac{\partial_r a}{a} - \frac{\partial_r b}{b} - \frac{2}{r} \right) \biggr] , \label{eq:sphere-Rrr}
\end{align}
\begin{align}
{}^{(3)}R = &- \frac{1}{a e^{4 \chi}} \biggl[ \frac{\partial^2_r a}{2a} + \frac{\partial^2_r b}{b} - a \partial_r \hD^r - \left( \frac{\partial_r a}{a} \right)^2  \nonumber \\
&+ \frac{1}{2} \left( \frac{\partial_r b}{b} \right)^2 + \frac{2}{rb} \left( 3 - \frac{a}{b} \right) \partial_r b + \frac{4}{r^2} \left( 1 - \frac{a}{b} \right) \nonumber \\
&+ 8 \left( \partial^2_r \chi + ( \partial_r \chi )^2 \right) - 8 \partial_r \chi \left( \frac{\partial_r a}{2a} - \frac{\partial_r b}{b} - \frac{2}{r} \right) \biggr] .
\label{eq:sphere-RSCAL}
\end{align}

The stress tensor of the minimally coupled scalar is
\begin{equation}
T_{\mu\nu} = \nabla_\mu\phi\,\nabla_\nu\phi - g_{\mu\nu}\!\left(\frac{1}{2}\nabla_\alpha\phi\,\nabla^\alpha\phi + V(\phi)\right) .
\end{equation}
Following Ref.~\cite{Alcubierre:2011pkc}, we define
\begin{align}
\Pi &\equiv n^\mu \nabla_\mu \phi = \frac{1}{\alpha}\!\left(\partial_t \phi - \beta^r \partial_r \phi\right) , \label{eq:defPi}\\
\Psi &\equiv \partial_r \phi ~, \label{eq:defPsi}
\end{align}
where $n^\mu$ is the future-directed unit normal to a spatial slice.
The matter sources in Eqs.~(\ref{eq:CCZ4_K})--(\ref{eq:CCZ4_Deltar}) are
\begin{align}
\rho_E &= \frac{1}{2}\Pi^2 + \frac{1}{2}\frac{e^{-4\chi}}{a}\Psi^2 + V(\phi)  ~, \label{eq:rhoE}\\
S_r &= -\Pi\,\Psi  ~, \\
S_a &= \frac{1}{2}\Pi^2 + \frac{1}{2}\frac{e^{-4\chi}}{a}\Psi^2 - V(\phi)  ~, \\
S_b &= \frac{1}{2}\Pi^2 - \frac{1}{2}\frac{e^{-4\chi}}{a}\Psi^2 - V(\phi) ~.
\end{align}
The variables $\{\phi,\Psi,\Pi\}$ evolve according to~\cite{Alcubierre:2011pkc}
\begin{align}
\partial_t \phi &= \beta^r \Psi + \alpha\,\Pi ~, \label{eq:evophi}\\
\partial_t \Psi &= \partial_r\!\left(\beta^r \Psi + \alpha\,\Pi\right) , \label{eq:evoPsi}\\
\partial_t \Pi &= \beta^r\partial_r\Pi + \frac{\alpha\, e^{-4\chi}}{a}\biggl[\partial_r\Psi + \Psi\!\left(\frac{2}{r} - \frac{\partial_r a}{2a} + \frac{\partial_r b}{b} + 2\,\partial_r\chi\right)\biggr] \nonumber \\
&\quad + \frac{e^{-4\chi}}{a}\,\Psi\,\partial_r\alpha + \alpha\, \left(\widehat K+2\Theta\right)\Pi - \alpha\,\frac{dV}{d\phi}  ~. \label{eq:evoPi}
\end{align}
Equation~\eqref{eq:evoPsi}, the radial derivative of Eq.~\eqref{eq:evophi}, is evolved independently to give a first-order reduction of the Klein--Gordon equation.

\subsection{Gauge conditions}

During inflation we use geodesic slicing, $\alpha=1$ and $\beta^r=0$.
For collapse the lapse obeys a cosmological Bona--Mass\'o condition,
\begin{equation}
\partial_t\alpha=-\mu_L\alpha^p\bigl(K_{\rm phys}-\langle K_{\rm phys}\rangle\bigr),
\end{equation}
where $K_{\rm phys}=\widehat K+2\Theta$.
Subtracting its radial average removes the homogeneous cosmological contribution to the lapse evolution.
We use $0.1\leq\mu_L\leq1$ and $1\leq p\leq2$, choosing values which maintain stable horizon-penetrating slices.

\subsection{Numerical implementation}

We integrate the BSSN equations with a fourth-order Runge--Kutta method~\cite{Joana:2024ltg,Kou:2019bbc}.
Fourth-order finite differences are used on a uniform cell-centred grid, with sixth-order Kreiss--Oliger dissipation to control grid-scale noise.
Each boundary has three ghost cells.

The domains extend to $r_{\mathrm{max}}=24$--$50\,k_\star^{-1}$ and contain $N=5000$--$10000$ radial cells.Parity at the origin is enforced by reflecting the ghost cells; at the outer boundary the fields are extrapolated towards the FLRW solution.

\subsection{Construction of initial data}

The continuous inflationary evolutions and the threshold scans use distinct initial-data constructions.

\subsubsection{Inflationary perturbations}

For all three inflationary configurations, the field profile~\eqref{eq:IC} is specified on a conformally flat slice with $a=b=1$ and $A_a=A_b=\hD^r=0$.
The window is unity for $r\leq2r_{\max}/3$, falls smoothly to zero over $0.2r_{\max}$, and vanishes thereafter.
It does not affect the central sinc profile.
We take the homogeneous SR momentum, $\overline\Pi=-V'(\overline\phi)/\sqrt{3V(\overline\phi)}$.
The momentum constraint then gives
\begin{equation}
K_{\rm phys}(r)=\overline K+\frac{3}{2}\overline\Pi\,[\phi(r)-\overline\phi],
\qquad
\overline K=-\sqrt{3\overline\rho_E}.
\end{equation}
Since $A_a=A_b=0$ on this slice, the Hamiltonian constraint takes the form
\begin{equation}
\mathcal{H} = {}^{(3)}R + \tfrac{2}{3}K_{\rm phys}^2 - 2\rho_E = 0. \label{eqn:HamSimp}
\end{equation}
Writing $\psi=e^\chi$ reduces this to a nonlinear ordinary differential equation, which we solve iteratively by relaxation.
The trapped, overdense, and void solutions in the main text are evolved continuously from these data, through the SR--USR--kination transition and into collapse.
The coordinates are not rescaled and the evolved fields are not reconstructed; only the gauge parameters change when horizon-penetrating slices become necessary.

\subsubsection{Extraction of the super-Hubble curvature profile}

For the threshold scans, the curvature on uniform-field slices is extracted from a checkpoint after the transition to kination.
The logarithmic local volume of the numerical spatial metric is
\begin{equation}
  \mathcal{N}_{\rm vol}(r)
  = \frac{1}{3}\ln\!\left(\frac{\sqrt{\gamma}}{\sqrt{\gamma_{\rm flat}}}\right)
  = 2\chi + \frac{1}{6}\ln a + \frac{1}{3}\ln b ,
\end{equation}
and its perturbation with respect to the asymptotic FLRW region is
\begin{equation}
  \zeta_{\rm vol}(r)=\mathcal{N}_{\rm vol}(r)-\overline{\mathcal{N}}_{\rm vol}.
\end{equation}
Because $\Pi=n^\mu\nabla_\mu\phi$ is the proper-time derivative along the slice normal, the associated expansion rate is
\begin{equation}
  H= -\frac{\widehat K+2\Theta}{3},
\end{equation}
with no additional lapse factor.
To first order in the field displacement and at leading order in gradients, the uniform-field curvature is
\begin{equation}
  \calR_{\rm lin}(r)=\zeta_{\rm vol}(r)
  -\overline{H}\,
   \frac{\phi(r)-\overline\phi}{\overline\Pi} .
  \label{eq:Rextract}
\end{equation}
The minus sign follows by cancelling a local time shift $\delta\tau$, under which $\delta\ln a=\overline H\delta\tau$ and $\delta\phi=\overline\Pi\delta\tau$.
Thus Eqs.~\eqref{eq:Rextract} and~\eqref{eq:RextractNL} describe the uniform-field (comoving) curvature at leading order in the gradient expansion.
For a finite displacement we use the separate-universe expression introduced in Eq.~\eqref{eq:Rdecomp},
\begin{equation}
  \calR(r)=\zeta_{\rm vol}(r)
  -\int_{\overline\phi}^{\phi(r)}
       \frac{\overline{H}(\varphi)}{\overline\Pi(\varphi)}\,d\varphi
  +\mathcal{O}\!\left(\frac{k_\star^2}{a^2H^2}\right).
  \label{eq:RextractNL}
\end{equation}
The integral follows the background phase-space trajectory, on which $\overline H$ and $\overline\Pi$ may be regarded as functions of $\phi$.
Since $(\overline H/\overline\Pi)d\varphi=\overline Hdt=d\overline{\calN}$, it is simply the background e-fold interval between $\overline\phi$ and $\phi(r)$.
In pure kination, $V\simeq0$ and $3\overline H^2=\overline\Pi^2/2$, so the ratio $\overline H/\overline\Pi$ becomes constant.
On our rolling branch $\overline\Pi<0$, and hence
\begin{equation}
  \calR(r)=\zeta_{\rm vol}(r)+\frac{\phi(r)-\overline\phi}{\sqrt{6}}.
\end{equation}
Consequently $\calR_{\rm lin}\simeq\calR$ at the kination checkpoints used for the threshold calculations.
Before exporting a profile we require $k_\star/(aH)\ll1$ and check that $|\sqrt6\,\overline H/\overline\Pi|\simeq1$.
The dimensionless profile supplied to the scan is
\begin{equation}
  P(k_\star r)=\frac{\calR(r)}{\calR(0)},\qquad P(0)=1.
\end{equation}

\begin{figure}[h]
    \centering
    \includegraphics[width=\columnwidth]{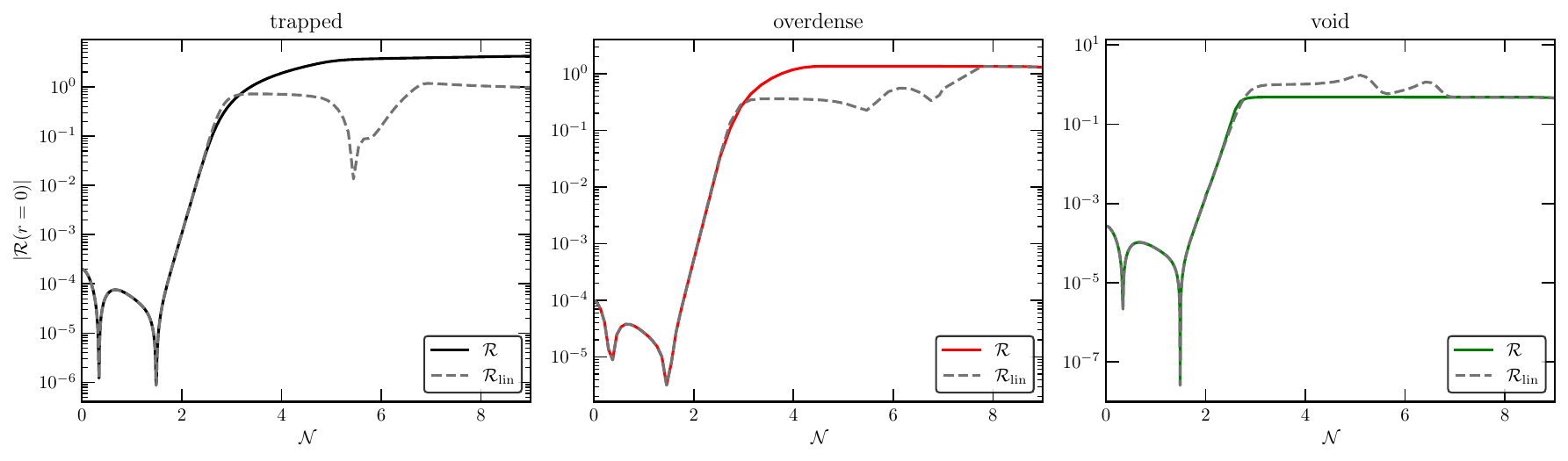}
    \caption{%
    Central comoving curvature for the trapped (left), overdense (centre), and void (right) solutions, evaluated with the separate-universe expression~\eqref{eq:RextractNL} (solid) and its linearisation~\eqref{eq:Rextract} (grey dashed).
    The former freezes after Hubble exit for the overdense and void cases and continues to grow for the trapped case.
    The linearised result has transient excursions during the USR period, while the linearisation regime fails, but becomes valid again and agrees with $\calR$ once kination is established.
    Curves end before Hubble re-entry, after which $\calR$ is no longer conserved.
    }
    \label{fig:Rcomparison}
\end{figure}

Figure~\ref{fig:Rcomparison} compares the two curvature estimates at the centre of each selected perturbation.
The separate-universe result becomes constant soon after Hubble exit for the overdense and void solutions, but grows monotonically for the trapped solution.
During the USR--kination transition, $\overline H/\overline\Pi$ varies by orders of magnitude across a finite field displacement and the linearised formula develops large transient excursions, due to the failure of the linearisation validity. 
After the USR/kination transition, the linear estimate becomes valid again. 

\subsubsection{Threshold scans}

Thresholds are obtained from independent kination simulations for two classes of profile.
The first consists of the overdense and void profiles extracted from the two selected initial configurations with Eq.~\eqref{eq:RextractNL} once they are conserved on super-Hubble scales.
We normalise them as $P(k_\star r)=\calR(r)/\calR(0)$ and scan the amplitude in $\calR(r)=\calR_0P(k_\star r)$.
The second class is the analytic reference $\calR(r)=\calR_0\mathrm{sinc}(k_\star r)$.
We then construct new, constraint-satisfying kination data.
We set $\chi=\calR/2$, $\phi=\overline\phi_\mathrm{bkg}$, $K_{\rm phys}=-3H_\mathrm{bkg}$, $a=b=1$, $A_a=0$, and $\Theta=0$; therefore $\widehat K=K_{\rm phys}$ initially.
The momentum constraint is identically satisfied because $\Psi=0$, $A_a=A_b=0$, and $K_{\rm phys}$ is spatially uniform. The Hamiltonian constraint~\eqref{eqn:HamSimp} then determines $\Pi$.

\subsection{Mass and apparent horizons}

\begin{figure}[h]
    \centering
    \includegraphics[width=\columnwidth]{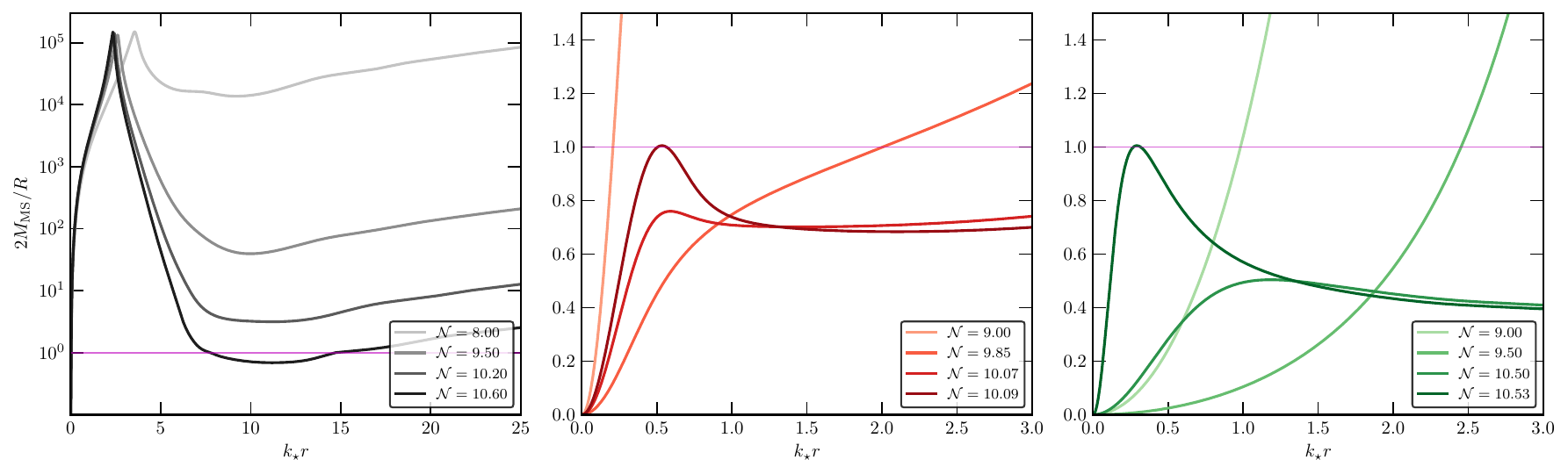}
    \caption{%
    Misner--Sharp compactness during collapse for the three PBH-forming channels.
    Darker curves denote later times; the magenta line at $2M_\mathrm{MS}/R=1$ marks a marginal surface.
    }
    \label{fig:collapse_app}
\end{figure}

Apparent horizons are found from the outgoing null expansion, whilst the Misner--Sharp compactness is used as a diagnostic.
In our spherical variables,
\begin{equation}
    \Theta_{\pm} = \pm\frac{e^{-2\chi}}{\sqrt{a}} \left( 4\, \partial_r \chi + \frac{2}{r} + \frac{ \partial_r b}{b} \right) + A_a - \frac{2}{3} K_{\rm phys} ~.
\end{equation}
The areal radius is $R=re^{2\chi}\sqrt b$, giving the Misner--Sharp mass
\begin{equation}
 M_{\mathrm{MS}}=\frac{R}{2}\left[1+\left(\frac{\partial_tR-\beta^r\partial_rR}{\alpha}\right)^2-\frac{e^{-4\chi}}{a}(\partial_rR)^2\right].
\end{equation}
Its derivatives are
\begin{align}
   \partial_t R &= e^{2\chi} r \sqrt{b} \left( 2 \,\partial_t \chi + \frac{1}{2} \frac{\partial_t b}{b} \right)  ~,\\
    \partial_r R & = e^{2\chi} \sqrt{b} \left(1 + 2 r\,\partial_r \chi
     + \frac{r}{2} \frac{\partial_r b}{b} \right) ~.
\end{align}
We take the apparent horizon to be the outermost root of $\Theta_+=0$ for which $\Theta_-<0$; the latter condition excludes cosmological turnaround surfaces.
There $2M_\mathrm{MS}/R=1$, and we report $M_\mathrm{BH}=M_\mathrm{MS}=R_\mathrm{AH}/(2G)=4\pi R_\mathrm{AH}$ in Figure~\ref{fig:masses}; Figure~\ref{fig:collapse_app} shows the compactness during collapse.

\subsection{Code validation}

We monitor the Hamiltonian constraint throughout each evolution.
With Z4c damping, its residual sources $\Theta$ through Eq.~\eqref{eq:CCZ4_Theta} and feeds back into $K$, $\chi$, and $\hD^r$ through the $\kappa_1$ terms.
For both Z4c and BSSN simulations we use the normalised residual
\begin{equation}
    \mathcal{H}_{\mathrm{rel}}  \equiv
    \frac{{}^{(3)}R-(A_a^2+2A_b^2)+\tfrac{2}{3}K_{\rm phys}^2-2\rho_E}
    {|{}^{(3)}R|+|A_a^2+2A_b^2|+\tfrac{2}{3}|K_{\rm phys}^2|+2|\rho_E|}.
    \label{eq:HamRel}
\end{equation}
We require at least $|\mathcal{H}_\mathrm{rel}|<10^{-2}$ throughout the physical domain outside the apparent horizon, which is largely satisfied in our simulations. 

\begin{figure}[h]
    \centering
    \includegraphics[width=\columnwidth]{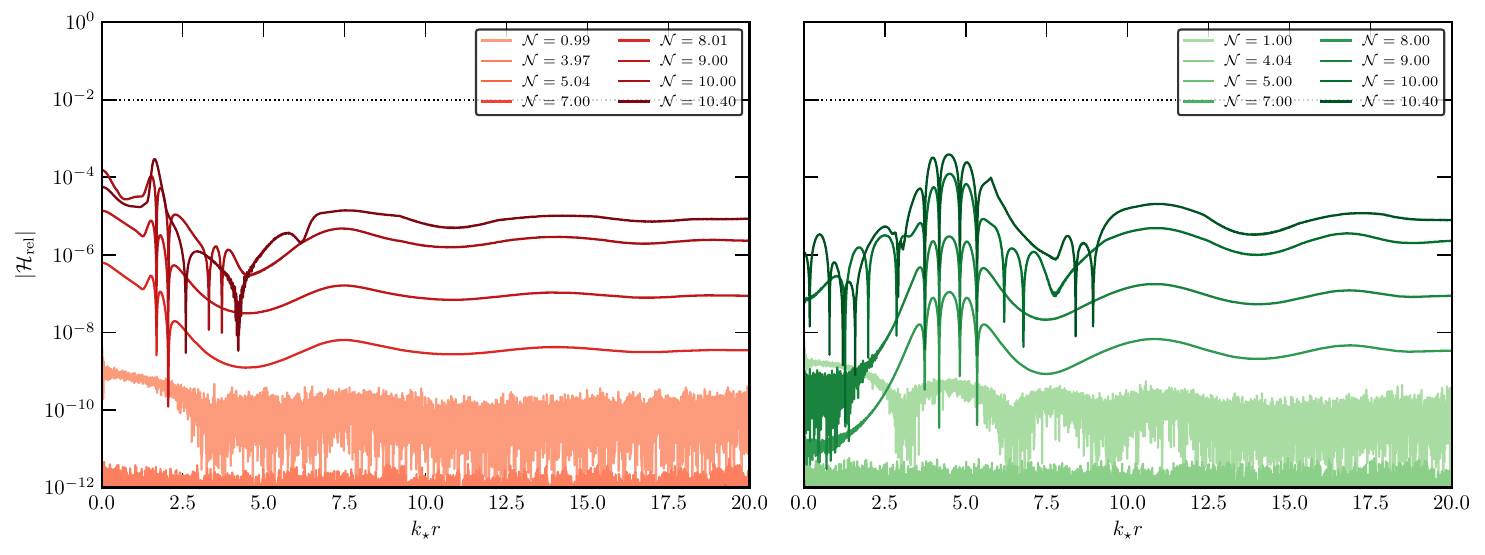}
    \caption{%
    Relative Hamiltonian-constraint residual~\eqref{eq:HamRel} for the selected overdense (left) and void (right) simulations at successive e-fold times.
    Darker curves denote later times; the dotted line is the adopted $10^{-2}$ bound.
    }
    \label{fig:Hrel_slices}
\end{figure}

Figure~\ref{fig:Hrel_slices} gives $|\mathcal{H}_{\rm rel}|$ for the selected overdense and void simulations at times spanning inflation, the transition to kination, re-entry, and collapse.
At the times shown it remains below $10^{-3}$, well within the adopted bound. 

\end{document}